# Acceso abierto en Argentina: una propuesta para el monitoreo de las publicaciones científicas con OpenAlex

*Carolina Unzurrunzaga (https://orcid.org/0000-0002-4383-0085)[1], Carolina Monti (https://orcid.org/0000-0002-8126-3712)[2], Gastón Zalba (https://orcid.org/0009-0003-3467-1363)[1], Juan Pablo Alperin (https://orcid.org/0000-0002-9344-7439)[3*]*

[1]*Universidad Nacional de La Plata, Argentina*
[2]*Consejo de Investigaciones Científicas (CONICET), Argentina*
[3] *Scholarly Communications Lab & School of Publishing, Simon Fraser University, Canada*
*\* autor correspondiente: juan@alperin.ca*

## Resumen

En este estudio se propone una metodología utilizando OpenAlex (OA) para monitorear el acceso abierto (AA) a las publicaciones científicas para el caso de Argentina, país donde rige el mandato de autoarchivo -Ley 26.899 (2013)-. Se conformó una muestra con 167.240 artículos de investigadores del Consejo Nacional de Investigaciones Científicas y Técnicas (CONICET) que se analizaron con técnicas estadísticas. Se estimó que OA puede representar entre 85-93% de los autores para todas las disciplinas, excepto Ciencias Sociales y Humanidades, donde solo alcanza al 47%. Se calculó que 41% de los artículos publicados entre 1953-2021 incluidos en la fuente están en AA, porcentaje que sube a 46% al considerar exclusivamente el periodo post ley (2014-2021). En ambos periodos es la vía dorada la que representa mayor proporción. Al comparar periodos iguales post y pre ley, se observó que la tendencia en alza de la vía dorada era preexistente a la legislación y la disponibilidad de artículos cerrados en repositorios aumentó un 5% a lo que se estima en base a tendencias existentes. Se concluye que si bien la vía verde ha tenido una evolución positiva, ha sido la publicación en revistas doradas lo que ha impulsado más rápidamente el acceso a la producción argentina. Asimismo, que la metodología basada en OA, piloteada aquí por primera vez, es viable para monitorear el AA en Argentina ya que arroja porcentajes similares a otros estudios nacionales e internacionales.

## Contribuciones

**Conceptualización:** Carolina Unzurrunzaga, Carolina Monti, and Juan P. Alperin.
**Curación de datos:** Carolina Unzurrunzaga, Carolina Monti, and Juan P. Alperin.
**Análisis Formal:** Carolina Unzurrunzaga, Carolina Monti, and Juan P. Alperin.
**Metodología:** Gastón Zalba and Juan P. Alperin. Software: Gastón Zalba and Juan P. Alperin.
**Supervision:** Juan P. Alperin.
**Validación:** Carolina Unzurrunzaga, Carolina Monti, and Juan P. Alperin.
**Visualization:** Juan P. Alperin.
**Redacción - borrador original:** Carolina Unzurrunzaga, Carolina Monti, and Juan P. Alperin.
**Escritura, revisión y edición:** Carolina Unzurrunzaga, Carolina Monti, Gastón Zalba, and Juan P. Alperin.

# Open access in Argentina: a proposal for tracking scientific publications with OpenAlex

*Carolina Unzurrunzaga (https://orcid.org/0000-0002-4383-0085)[1], Carolina Monti (https://orcid.org/0000-0002-8126-3712)[2], Gastón Zalba (https://orcid.org/0009-0003-3467-1363)[1], Juan Pablo Alperin (https://orcid.org/0000-0002-9344-7439)[3*]*

[1]*Universidad Nacional de La Plata, Argentina*
[2]*Consejo de Investigaciones Científicas (CONICET), Argentina*
[3] *Scholarly Communications Lab & School of Publishing, Simon Fraser University, Canada*
\* *autor correspondiente: juan@alperin.ca*

## Abstract

This study proposes a methodology using OpenAlex (OA) for tracking Open Access publications in the case of Argentina, a country where a self-archiving mandate has been in effect since 2013 ( Law 26.899, 2013). A sample of 167,240 papers by researchers from the National Council for Scientific and Technical Research (CONICET) was created and analyzed using statistical techniques. We estimate that OA is able to capture between 85-93% of authors for all disciplines, with the exception of Social Sciences and Humanities, where it only reaches an estimated 47%. The availability of papers in Open Access was calculated to be 41% for the period 1953-2021 and 46% when considering exclusively the post-law period (2014-2021). In both periods, gold Open Access made up the most common route. When comparing equal periods post and pre-law, we observed that the upward trend of gold Open Access was pre-existing to the legislation and the availability of closed articles in repositories increased by 5% to what is estimated based on existing trends. However, while the green route has had a positive evolution, it has been the publication in gold journals that has boosted access to Argentine production more rapidly. We concluded that the OA-based methodology, piloted here for the first time, is viable for tracking Open Access in Argentina since it yields percentages similar to other national and international studies.

## Keywords

*Open access; Open access policy; Mandate; Methodology; Open Access Monitoring; CONICET; Argentina; Scientific publications; OpenAlex*

## Author Contributions

Conceptualization: Carolina Unzurrunzaga, Carolina Monti, and Juan P. Alperin. Data curation: Carolina Unzurrunzaga, Carolina Monti, and Juan P. Alperin. Formal analysis: Carolina Unzurrunzaga, Carolina Monti, and Juan P. Alperin. Methodology: Gastón Zalba and Juan P. Alperin. Software: Gastón Zalba and Juan P. Alperin. Supervision: Juan P. Alperin. Validation: Carolina Unzurrunzaga, Carolina Monti, and Juan P. Alperin. Visualization: Juan P. Alperin. Writing - original draft: Carolina Unzurrunzaga, Carolina Monti, and Juan P. Alperin. Writing - review & editing: Carolina Unzurrunzaga, Carolina Monti, Gastón Zalba, and Juan P. Alperin.

## 1. Introducción

En las últimas décadas cada vez son más los financiadores de la ciencia que exigen que los resultados de las investigaciones que solventan parcial o totalmente, estén disponibles bajo la modalidad de acceso abierto. Países e instituciones de todo el mundo han ido adoptado políticas, desarrollando infraestructura e instrumentos para que el conocimiento científico logre mayor circulación y alcance.

Por su parte las editoriales comerciales internacionales han ido adaptando sus modelos de negocio a los requerimientos de los financiadores y complejizando la concreción del acceso abierto al conocimiento. En Europa, por ejemplo, los llamados acuerdos transformativos han ganado terreno de la mano del Plan S validando altos costos por procesamiento de artículos (APC) y el pago por el acceso. Los onerosos APC que cobran revistas de acceso abierto o híbridas están restringiendo las posibilidades de muchos autores al momento de elegir una publicación y generan aún más inequidades en la comunicación de la ciencia (Debat y Babini, 2020; BOAI, 2022).

Determinar cuánto se ha avanzado en concretar las políticas de acceso abierto parecería una tarea sencilla de realizar si se contara con las herramientas necesarias, sin embargo, la información actualizada disponible es escasa. Nuevos estudios han mostrado cómo ha sido el avance del acceso abierto de manera global (Piwowar et al., 2018, Martín-Martín et al., 2018, Huang et al., 2020, Robinson-Garcia, Costa, van Leeuwen, 2020), como está la situación en determinados países (Alemania, Hobert et al., 2021; Finlandia, Pölönen et al., 2020; Francia, Jeangirard, 2019; Cataluña, Rovira, Urbano, Abadal, 2019, entre otros), regiones (Latinoamérica y Caribe, Minniti, Santoro y Belli, 2018; European Commission, Directorate-General for Research and Innovation, 2021) o a nivel de instituciones (Uribe et al 2019; Bernal y Román Molina, 2022). Incluso, se ha observado que la existencia de políticas de mandato junto con las de seguimiento y monitoreo repercuten en mejores tasas de publicación en acceso abierto (Larivière y Sugimoto, 2018; Huang et al., 2020); también, cómo con la presencia de financiamiento europeo aumenta la probabilidad de contar con publicaciones abiertas en determinadas disciplinas (Morillo, 2020).

América Latina es reconocida internacionalmente por sus iniciativas de acceso abierto y un fuerte número de revistas que no cobran a los autores ni a los lectores y suelen ser financiadas por instituciones o asociaciones de ciencia y tecnología, modelo ahora denominado diamante. No obstante, son pocos los países que cuentan con una política de acceso abierto nacional aprobada por vía legislativa que imponga el mandato de autoarchivo, entre estos: Perú (2013), Argentina (2013) y México (2014)[1]. Aunque son numerosas las iniciativas que se desarrollan desde organismos de ciencia y tecnología de toda la región, tal como sucede en el caso de Brasil (Babini y Rovelli, 2020; Salatino y Banzato, 2021). Hasta el momento y a pesar de los esfuerzos de distintos actores, las estadísticas regionales[2] y nacionales no reflejan datos que permitan contemplar los avances del acceso abierto y la política

---

[1] Se conocen además casos de otros países de la región como por ejemplo, Colombia que aprobó en 2019 un conjunto de directrices sobre ciencia abierta denominadas "Lineamientos para una política de ciencia abierta en Colombia" (Res. N° 0167); y Brasil, que a pesar de apoyar el movimiento de acceso abierto desde sus inicios y contar con varias iniciativas importantes como la Declaración de Salvador (2005), la Carta de São Paulo (2005) y el Manifiesto Brasileño de Apoyo al Acceso Libre a la Información Científica (2005) no ha logrado aún la sanción de una ley (Babini y Rovelli, 2020).

[2] Por ejemplo, para América Latina las estadísticas publicadas por la Red de Indicadores de Ciencia y Tecnología - Iberoamericana e Interamericana (RICYT) (http://www.ricyt.org/category/indicadores/), avalada por la OEI y la Unesco con aportes de los ministerios de ciencia y tecnología de la mayoría de los países iberoamericanos, lleva conteos de indicadores de

emprendida. En general, siguiendo la tendencia internacional, solo contabilizan publicaciones indizadas en el *mainstream* sin hacer referencia a su modalidad de acceso e invisibilizando la cuantiosa producción publicada en revistas de la región (Beigel et al. 2022).

En este artículo se propone una metodología de estudio que utiliza *OpenAlex*, base de datos abierta y gratuita poco explorada hasta el momento, para monitorear las publicaciones científicas de Argentina y determinar sus formas de acceso. Se espera que ésta aporte al desarrollo de procedimientos propios en distintas instituciones y países que permita monitorear el avance del acceso abierto utilizando fuentes más comprensivas que las utilizadas masivamente como WoS y Scopus. Para testear la metodología y la herramienta desarrollada se trabajó específicamente tomando como caso al Consejo Nacional de Investigaciones Científicas y Técnicas (CONICET) de Argentina. Con los datos recogidos, se buscó determinar cuánto de la producción generada en esta institución está disponible en acceso abierto, analizar el efecto del mandato establecido mediante la Ley nacional 26.899 (2013) que impone el depósito en repositorios institucionales propios -o compartidos- y detectar áreas disciplinares en donde podría haber dificultades para cumplirlo.

## 2. Acerca de las base de datos científicas, sus usos y cobertura

A lo largo de la historia, los servicios de información científica y sus bases de datos se constituyeron como fuentes para la realización de estudios que buscan conocer, describir, caracterizar y analizar las publicaciones científicas de diferentes dominios. Estas bases, además de tener diferentes objetivos, disponen sus datos de forma abierta o bajo suscripción, cubren distintas disciplinas, realizan o no algún tipo de selección y ofrecen indicadores bibliométricos que se calculan con base en sus colecciones (Vuotto et al., 2020).

Tradicionalmente las bases de datos Web of Science (WoS) y Scopus por el reconocimiento que tienen en el campo científico y los índices de citación que ofrecen, han sido empleadas por innumerables investigaciones para describir la productividad y medir el impacto en los distintos dominios. Incluso, y a pesar de las críticas recibidas por su falta de cobertura (Rozemblum et al., 2021), son utilizadas como fuente para el cálculo de indicadores oficiales de Ciencia y Tecnología en varios países y regiones dejando de lado otras fuentes que reúnen y visibilizan la producción de manera regional.

En las últimas décadas se desarrollaron un número importante de servicios de información que se caracterizan por agregar otros servicios conformando así corpus amplios de literatura científica. Entre los que ofrecen una versión gratuita pueden mencionarse Lens (2000, inicialmente solo de patentes), Google Scholar -GS- (2004), Microsoft Academic Graph -MAG- (2016, discontinuado en diciembre de 2021), Crossref (2017) y Dimensions (2018). Estudios comparados han demostrado que, por ejemplo, GS o Dimensions tienen una mejor cobertura que WoS o Scopus (Harzing, 2019; Delgado López-Cózar et al., 2019); mientras que MAG es más abarcativa en comparación con WoS y Scopus sobre todo en Sociales, Humanidades y Artes, que son las disciplinas menos representadas en ambas bases (Huang et al., 2020).

producción que consideran únicamente artículos indizados en base de datos comerciales, en concreto, Scopus y Science Citation Index sin considerar revistas incluidas en servicios regionales de larga trayectoria como son Latindex, SciELO y RedALyC.

Para determinar la disponibilidad en acceso abierto, si bien han existido otros desarrollos particulares, Unpaywall se ha convertido en la fuente más utilizada para realizar estudios que posibiliten su monitoreo (Borrego, 2022). Ha sido comprobado que esta base de datos desarrollada por *OurResarch* -organización sin ánimo de lucro dedicada a los principios académicos- permite conocer el tipo de acceso de los materiales con una precisión muy alta (Piwowar et al. 2018) aunque Jeangirard (2019) al analizar su efectividad halló que para trabajos del periodo 2013-2017 hay entre un 3% y un 11% de falsos cerrados, es decir, artículos clasificados como cerrados que en realidad son de acceso abierto. Asimismo, se pudo observar que se suele combinar con el *Directory of Open Access Journals* (DOAJ), que provee información de las políticas de las revistas, a fin de calcular los costos de APC.

En este estudio, como se mencionó anteriormente, se utiliza la base de datos OpenAlex -en adelante OA- , que fue creada también por *OurResarch*. Durante el periodo estudiado, OpenAlex estaba en su versión "beta". Este servicio recopila trabajos científicos de diferentes fuentes: MAG, Crossref, Unpaywall (que a su vez incluye WoS y Scopus), ORCID, ROR, DOAJ, Pubmed, Pubmed Central, *The ISSN International Centre*, *Internet Archive*, repositorios como arXiv y Zenodo, entre otros. Sus fuentes de metadatos académicos son totalmente abiertas (datos 100% abiertos, API abierta para extracción de datos, código fuente abierto) (Priem, Piwowar y Orr, 2022). Además, el conjunto de datos OA ofrece la posibilidad de hacer búsquedas por cinco tipos de entidades académicas: trabajos científicos, autores, instituciones, fuentes (revistas, congresos, repositores, etc.) y conceptos[3].

Varios estudios recientes sobre la cobertura OA, WoS y Scopus encuentran que herramientas como OA permiten eludir la falta de cobertura de las bases de datos más selectivas (Cubert et al., 2024; Jiao et al., 2023). Estar presente en estas dos últimas bases de datos suele considerarse como un indicador de calidad de las revistas académicas en la evaluación de la investigación, lo que obstruye una visión más completa de un sistema de investigación científica. Simard y otros (2024) demostraron que la mayoría (60%) de las revistas incluidas en DOAJ también lo están en OA, mientras que WoS y Scopus indexan menos de la mitad. Particularmente, en WoS encontraron que la presencia de las revistas diamante es baja (Simard et al., 2024).

## 3. El caso de Argentina: avances para el acceso abierto

Argentina es uno de los países latinoamericanos en donde se desarrollaron tempranamente políticas de acceso abierto a nivel nacional. El trabajo de distintos actores comenzó a forjar desde 2009 un camino que apuntó una política inclinada hacia la reunión y difusión de la producción científica-tecnológica nacional en repositorios institucionales. Las políticas guiadas por el Ministerio de Ciencia, Tecnología e Innovación de la Nación (MinCyT) tuvieron su punto cúlmine en diciembre de 2013 cuando se promulgó la Ley 26.899 que instituyó el mandato de depósito en repositorios nacionales para investigaciones financiadas con fondos públicos (Fushimi et al., 2021). La misma, que entró efectivamente en vigencia en 2014, estableció en su artículo 5° que

> Los investigadores, tecnólogos, docentes, becarios de posdoctorado y estudiantes de maestría y doctorado cuya actividad de investigación sea financiada con fondos públicos, deberán depositar o autorizar expresamente el depósito de una copia de la versión final de su producción científico-tecnológica publicada o aceptada para publicación y/o que haya atravesado un proceso de aprobación por una autoridad competente o con jurisdicción en la materia, en los repositorios digitales de acceso abierto de sus instituciones, en un plazo no mayor a los seis (6) meses desde la fecha de su publicación oficial o de su aprobación.

---

[3] A partir del 2023 la plataforma agregó nuevas entidades: trabajos científicos, autores, instituciones, fuentes, conceptos, editoriales, financiadores y localización.

Siguiendo con está línea, desde el MinCyT se conformó en 2021 el Comité Asesor en Ciencia Abierta y Ciudadana que realizó un diagnóstico de situación en el país y propuso líneas de acción para fortalecer los repositorios y respaldar prácticas académicas e investigativas que favorezcan la apertura. En el informe realizado por el Comité, no se da cuenta de la existencia de una iniciativa de monitoreo a nivel país que indique el porcentaje de disponibilidad en abierto de la producción nacional (Comité Asesor en Ciencia Abierta y Ciudadana, 2022).

Respecto a la disponibilidad de la producción argentina en acceso abierto previo a la sanción de la ley, Miguel, Gómez y Bongiovani (2012) mostraban un panorama alentador al analizar las publicaciones incluidas en Scopus: si bien sólo el 25% de los artículos estaban publicados en revistas de acceso abierto, encontraron que otro 44% podrían ser accesibles bajo esta modalidad al considerar las políticas de autoarchivo de las revistas en que publicaban los investigadores de medicina, física y astronomía, agricultura y ciencias biológicas, y ciencias sociales y humanidades. Posterior a la ley, las mismas autoras identificaron que el 61% de la producción de ciencias sociales incluida en Scopus publicada en 2017 estaba disponible en acceso abierto, mostrando un incremento del 18% respecto al estudio anterior, en el cual, para esta área específica habían identificado un 43% de acceso abierto real, lo cual explican podría deberse al aumento de revistas latinoamericanas en la base de datos (Bongiovani y Miguel, 2019).

Recientemente, un informe a nivel nacional que se basó en 134.412 publicaciones realizadas entre 2013 y 2020 que fueron incluidas en distintas bases de datos (WoS, Scopus, Unpaywall, Google Scholar, Lens y DOAJ) mostró que un 54,4% (73.271) de artículos publicados por investigadores con afiliación argentina estaba disponible en acceso abierto mientras que estimó altos desembolsos realizados para el pago de APC. Asimismo, al desagregar un poco más de 16 mil trabajos con filiación del CONICET, se encontró que la vía dorada -con y sin APC- era la principal opción de los autores mostrando esta un crecimiento notorio en los últimos años (Vélez Cuartas et al., 2022).

A diferencia de otros trabajos mencionados, aquí proponemos una metodología que parte de un conjunto conocido de individuos —en nuestro caso, aquellos afiliados con el CONICET al momento de la extracción de los datos—. Se considera que de esta forma, podría ser replicada para cualquier lista de individuos —por ejemplo, aquellos afiliados con un centro de investigación o una universidad—, para brindar un panorama de sus trayectorias y conocer con mayor profundidad el impacto de los mandatos de acceso abierto, sin limitarse a la falta de datos o cambios en la afiliación de los autores. Esto va en concordancia con otros estudios recientes que proponen la idea de no trabajar con un universo de artículos sino con un universo de investigadores (Beigel et al., 2023).

## 4. Metodología

Para realizar un monitoreo del acceso abierto, se diseñó una metodología cuantitativa que prevé la obtención de información de dos fuentes: el sistema de currículum vitae (CV) de los investigadores, para el caso concreto la información de los nombres y la producción informada que se publica en el sitio web del CONICET, y OA. Se eligió esta última entre otras fuentes por tratarse de una herramienta hasta el momento poco explorada que reúne, como se ha mencionado, los corpus de datos de los sistemas de información científica más variados contando, además, con datos sobre la disponibilidad en acceso abierto sin requerir combinar otras bases de datos.

### *4.1. Elección del caso*

Se optó por trabajar con Argentina que, como se exponía, es en la región de América Latina uno de los primeros países en donde se impuso el mandato a los investigadores que reciben financiamiento del Estado. Por la falta de set de datos actualizados de la producción científica a nivel nacional[4], se decidió avanzar en conocer la situación del CONICET, de cuya web se puede obtener información de los CVs de los investigadores tanto académica (escalafón, región geográfica, disciplina científica) como de publicaciones informadas en su Sistema Integral de Gestión y Evaluación (SIGEVA), sistema que contiene el banco de currículum. Este último aspecto resultó fundamental para tener una aproximación a la representatividad de la fuente OA.

En Argentina el CONICET se constituye como el principal organismo de ciencia y tecnología del país, por la cantidad de ingresos que recibe y la cantidad de personal con dedicación exclusiva que lo integra. Además ha sido un actor importante tras promulgarse la ley nacional de acceso abierto con la creación de su repositorio CONICET Digital en 2015, el de mayor tamaño a nivel nacional[5] (Zanotti, Isoglio y Piccotto, 2021). Por la estructura de la organización, su dimensión y su distribución en el territorio nacional, la institución sustrae de SIGEVA los documentos y los datos de los mismos que los investigadores informan en el organismo a través de esa plataforma y luego son curados por personal que conforma la Red Federal de Especialistas de CONICET Digital con lugar de trabajo en las distintas unidades de investigación para ser incorporados al repositorio.

En cuanto a las publicaciones de investigadores del CONICET, se cuenta con un estudio reciente de Beigel y Gallardo (2021) realizado sobre la producción informada en sus CVs a inicios de 2020 que permite tener algunas características generales del grupo, entre ellas: que el formato artículo representa la modalidad de publicación más frecuente y que los investigadores publicaron al menos un artículo, a razón de 32 en promedio (moda de 12). Asimismo, la productividad en términos de artículos es significativamente más baja para las mujeres quienes publicaron en promedio 28 artículos frente a 37 de los varones. También que el promedio más alto de artículos por persona corresponde a las Ciencias Exactas y Naturales (37), seguidas por las Ciencias Biológicas y de la Salud (34). Ciencias Sociales y Humanidades y Ciencias Agrarias, de las Ingenierías y de los materiales aparecen con valores un poco más bajos, 29 y 28 respectivamente.

### *4.2. Extracción de trabajos de OpenAlex*

Para poner en marcha la metodología se procedió del siguiente modo. En primer lugar, se trabajó en la obtención del listado de nombres de las personas que trabajan en el CONICET. Se extrajo información a través de un programa desarrollado *ad hoc* con técnica de *data scraping* de los investigadores registrados en el sitio web del organismo que pertenecen a un escalafón de la carrera de investigador (CIC) y la cantidad de publicaciones desagregadas por tipo documental informadas[6]. El universo de estudio quedó conformado por 12.292 personas para los cuales se contabilizaron unos 435.013 artículos que informaron al momento de extraer los datos (agosto de 2022).

---

[4] Los datos disponibles en el portal del Sistema de Información de Ciencia y Tecnología Argentino (https://sicytar.mincyt.gob.ar/) tiene información de las publicaciones hasta el 2017 y los datos del personal son de 2015 (https://sicytar.mincyt.gob.ar/estadisticas/#/b). En tanto, en el sitio de datos abiertos del Estado se encuentra publicada información hasta 2018 (https://www.datos.gob.ar/dataset/mincyt-personal-ciencia-tecnologia) (consultas realizadas en agosto 2023).
[5] Al 24/08/2023 el repositorio cuenta con 202.202 documentos, de los cuales 151.278 están disponibles en acceso abierto.

En segundo término, se procedió a la extracción de los trabajos disponibles en la base de datos OA a través de un *script* en *Python*[7] (Zalba, 2023) de instalación y ejecución local, que permite en base de un listado de nombres de personas extraer los datos de las publicaciones utilizando la API online de OA[8]. El *script* funciona armando variaciones de los nombres -con y sin tilde, iniciales de nombre, etc.- y realiza múltiples peticiones a la API que vuelven a ser filtradas descartando homónimos de otros países a partir de las instituciones declaradas en las afiliaciones (parámetro *country_code*). Permite determinar un porcentaje mínimo de trabajos del autor matcheado para considerarlo perteneciente al país seleccionado (parámetro *match_percentage*)[9]. Solo si un autor pasa estas comprobaciones sus trabajos son tomados como válidos y agregados a la base final.

Debido a que OA se encontraba en estado beta al momento de la realización de este estudio, se debieron sortear algunos inconvenientes al emplear el listado de nombres de personas que trabajan en el CONICET, por ejemplo, en ocasiones respondía de manera distinta a una misma petición y requirió de mayor desarrollo del código al buscar coincidencia en los nombres de autor por particularidades del idioma español (por ej. tildes y ñ) o cuestiones típicas de los nombres (ej. apellidos compuestos, con preposición y apellidos de casadas) (ver detalle en anexo 1). Asimismo, se detectó la existencia de varias entradas que resultaban válidas para un mismo autor entremezcladas con otras incorrectas, lo cual posiblemente pueda deberse a la multiplicidad de fuentes que integran OA. En conjunto con estos, se sumó la dificultad de la falta de tildes en nombres y apellidos en el *input*, por lo que se requirió el armado de un listado de nombres y apellidos que se acentúan con frecuencia para agregarlos al *script*. Entre los inconvenientes que no pudieron ser salvados deben mencionarse: la existencia de autores homónimos (o con nombres similares) y los nombres variantes, entre ellos, el uso indistinto de apellidos de casada y/o soltera, pseudónimos y firmas que utilizan iniciales.

Como tercer paso, se realizó el depuramiento de los trabajos extraídos. A falta de un estudio para evaluar la calidad de los datos de OA, se puso el foco en definir un umbral que permitiera reducir los errores de la extracción en cuanto a personas que presentaban una diferencia grande entre en el número de trabajos declarados en los CVs del organismo y los recuperados de OA. Para este fin, se consideró trabajar con la muestra de investigadores para los cuales esta diferencia no supera el 50% de los trabajos declarados. El corpus final quedó conformado por 280.011 trabajos publicados por las 12.292 personas, contando cada trabajo una vez por cada co-autor. Esto corresponde a 167.240 trabajos únicos, los cuales fueron analizados con técnicas de estadística descriptiva y bivariada (regresión lineal).

---

[6] Como *input* se usaron búsquedas por escalafón de la información disponible en el buscador de personas del organismo: https://www.conicet.gov.ar/new_scp/advancedsearch.php

[7] La aplicación y su código está disponible en el repositorio GitHub donde se puede encontrar información acerca de los requerimientos para instalarla, su guía de uso y configuración: https://github.com/GastonZalba/openalex-get

[8] La API de Open Alex puede consultarse en https://docs.openalex.org/

[9] Cabe aclarar que, como limitación, se han encontrado casos donde el país no está presente por lo que no se puede hacer esta comprobación.

En cuanto a las variables que se trabajaron, se utilizó la variable estado (*open_access.oa_status*) que es tomada por OA de la base de datos de Unpaywall para determinar la disponibilidad actual en acceso abierto. Asimismo, se empleó la variable *host_venue.url* a fin de identificar los repositorios argentinos.

## 5. Resultados

A continuación, se exponen en primera instancia los resultados de la comparación de los artículos informados versus los que conforman la muestra a fin de estimar la representatividad de OA para el caso de estudio. Luego, se da cuenta del estado en que los artículos publicados por los investigadores del CONICET se encuentran disponibles para su acceso. Posteriormente, para conocer los efectos del mandato establecido mediante la Ley nacional 26.899 (2013) se analizaron los resultados considerando tres muestras: una general que corresponde a trabajos publicados en el periodo 1953-2021 y otros dos recortes temporales: 2014-2021, considerando los 8 años posteriores a la promulgación de la norma y 2006-2021, para obtener un periodo comparable con los 8 años anteriores a la legislación.

### *5.1. Estimación de la representatividad de la fuente*

Para estimar la representatividad de OA respecto de la producción informada por los investigadores se comparó el total de artículos informado por cada investigador contra la cantidad extraída de la fuente y aquellos que conforman la muestra diseñada (tabla 1). Si bien el número de artículos recuperados llega al 80,3% de los informados, tras eliminar a los autores para los cuales había una gran discrepancia entre el número de artículos informados y los recuperados, se llegó a una muestra que cubriría un 64% de los trabajos informados y un 72% de los investigadores en la base original.

**Tabla 1.** Comparación por área temática del CONICET de artículos informados, recuperados e incluidos

| Área | Investigadores informados | Artículos informados | Investigadores recuperados | % Investigadores recuperados | Artículos recuperados | % Artículos recuperados | Investigadores incluidos | % Investigadores incluidos | Artículos incluidos | % Artículos incluidos |
|---|---|---|---|---|---|---|---|---|---|---|
| Cs. Agrarias, de las Ingenierías y de los materiales | 3260 | 100066 | 3218 | 98,7% | 84990 | 84,9% | 2608.0 | 80% | 71062 | 71% |
| Cs. Biológicas y de la Salud | 3544 | 137815 | 3511 | 99,1% | 123650 | 89,7% | 3037.0 | 85,7% | 106240 | 77,1% |
| Cs. Exactas y Naturales | 2638 | 104620 | 2611 | 99,0% | 97398 | 93,1% | 2182.0 | 82,7% | 79877 | 76,3% |
| Cs. Sociales y Humanidades | 2833 | 92107 | 2764 | 97,6% | 42806 | 46,5% | 995.0 | 35,1% | 22643 | 24,6% |
| Sin especificar | 17 | 405 | 16 | 94,1% | 316 | 78.0% | 8.0 | 41,1% | 189 | 46,7% |
| **Total** | **12292** | **435013** | **12120** | **98,6%** | **349160** | **80,3%** | **8830.0** | **71,8%** | **280011** | **64,4%** |

Fuente: Base de datos propia con información disponible en el sitio web oficial de CONICET (agosto 2022) y OpenAlex (septiembre 2022). *Nota:* El corpus final de artículos quedó conformado por 280.011 trabajos contando cada trabajo una vez por cada co-autor. Esto corresponde a 167.240 trabajos únicos, los cuales fueron analizados.

En líneas generales puede decirse que la representación de los trabajos de los investigadores recuperados en todas las grandes áreas temáticas es importante, considerando que superan el porcentaje general, variando entre el 85-93%, excepto para las Cs. Sociales y Humanidades donde la cobertura solo alcanza al 46,5% y desciende entre 10% y 20% en todas las disciplinas en la muestra analizada. Esto sucede a pesar de que OA incorpora portales de revistas y servicios como La Referencia, SciELO o RedALyC.

### 5.2. Disponibilidad en acceso abierto

Para el total de los 167.240 trabajos únicos extraídos, que abarcan el periodo 1953-2021, se ha podido identificar que 68.348 (41%) están disponibles en acceso abierto: un 16% (27,085) son dorados -artículos de libre lectura publicados en revistas de acceso abierto-; un 13% (21.068) son verdes -artículos publicados en revistas de acceso de pago que se han archivado en un repositorio-; 9% (14.813) son bronce -artículos de libre lectura en el sitio web del editor sin una licencia clara que otorgue ningún otro derecho de explotación-; y 3% (5.382) son híbridos -artículos publicados en revistas por suscripción que son de libre lectura desde el momento de su publicación con una licencia abierta gracias al pago realizado por el autor-. De los 15.210 artículos que no han podido determinarse su tipo de acceso, 13.073 no tienen DOI, lo que podría explicar la falta de datos de disponibilidad.

Un recorte del periodo 2006-2021 (Gráfico 1) nos permite contrastar los ocho años anteriores y posteriores a la ley. De esta manera puede observarse que la cantidad de trabajos accesibles desde la vía dorada tienen una tendencia a la suba incluso anterior a la aprobación de la ley. Esta categoría incluso es la que tiene los mayores cambios, pasando de representar sólo un 5% de los trabajos en 2006, a un 19% el año de la ley, y alcanzando un pico de 31% en 2020. En tanto, la publicación en revistas por suscripción bajo la modalidad híbrida que ofrecen la posibilidad de apertura mediante pago se mantienen relativamente estables, con fluctuaciones del 3% en 2006 al 4% en 2021. Por su parte, los artículos publicados en la vía bronce, se mantienen con valores similares en toda la serie mostrando cierta tendencia a la baja en los últimos años.

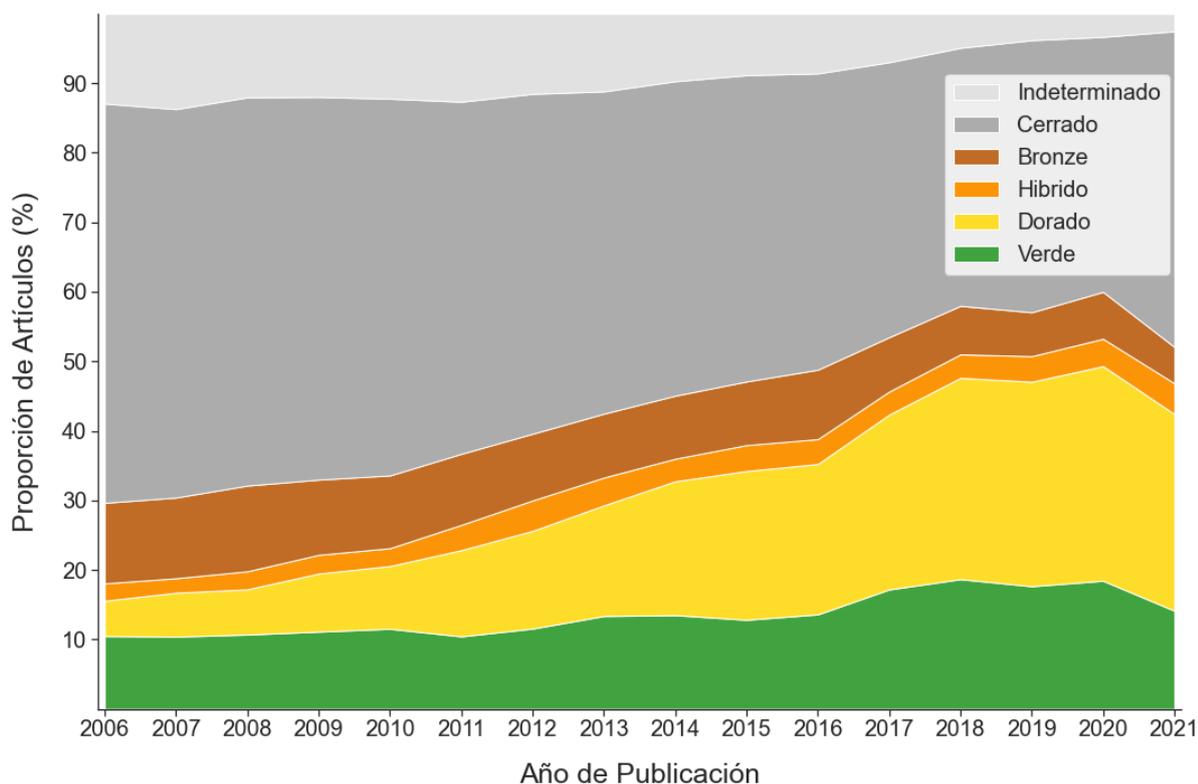

**Gráfico 1.** Tipo de acceso de los artículos encontrados para investigadores del CONICET (2006-2021)

Fuente: Base de datos propia basada en datos extraídos de Open Alex en agosto de 2022 (N = 127,705).

Asimismo, puede verse como la cantidad de trabajos solo disponibles a través de la vía verde muestra un leve aumento desde los años de creación de los primeros repositorios (De Volder, 2008; Fushimi et al., 2021) y presenta su mayor pico en 2018. Sin embargo, debe advertirse que en el gráfico 1 no es posible ver el crecimiento de la vía verde en general, ya que artículos publicados en acceso abierto directo (dorado o híbrido) podrían estar disponibles también en los repositorios, en este análisis cuentan los artículos cerrados presentes en repositorios.

Para comprender mejor la adopción del uso de repositorios, se analizó la proporción de trabajos depositados en repositorios argentinos e internacionales. En el gráfico 2 se pueden visualizar algunos picos que indican aumentos importantes: el primero en 2013 año de creación de la ley (pero antes a su implementación) y el segundo en 2017, año posterior a la implementación de la ley, en el cual se pusieron en marcha líneas de financiamiento para la creación y fortalecimiento de repositorios de instituciones de CyT. En tanto, la reducción de trabajos disponibles en repositorios que se observa en el 2021 podría deberse a la demora en la ingesta por el autoarchivo mediado a través del sistema de CV (Sigeva) del repositorio CONICET Digital.

**Gráfico 2.** Proporción de artículos depositados en repositorios argentinos e internacionales (2006-2021)

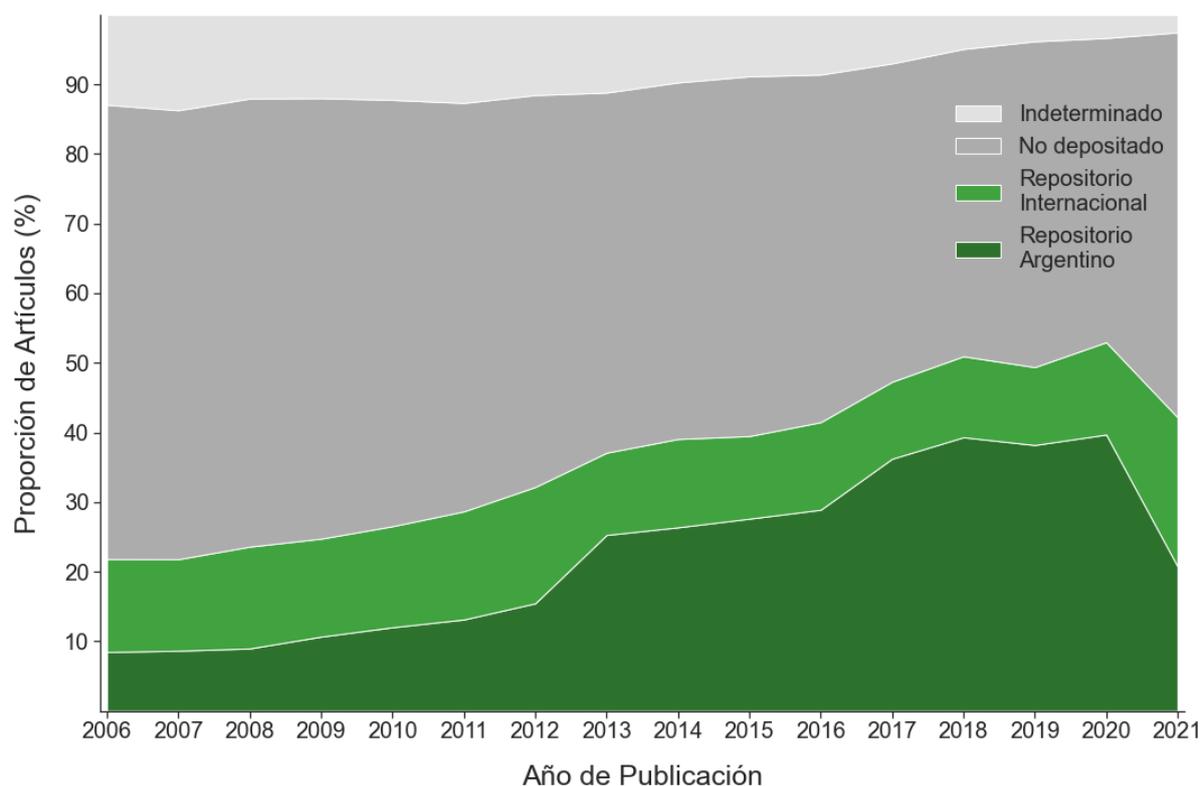

Fuente: Base de datos propia basada en datos extraídos de Open Alex en agosto de 2022 (N = 127,705).

### 5.3. Impacto de la ley de AA

Es claro que hay una tendencia hacia la apertura de los artículos publicados por investigadores del CONICET (Gráfico 1) y en la proporción de artículos que son depositados en repositorios (Gráfico 2). El efecto acumulativo de estas tendencias se puede ver al comparar la proporción de artículos abiertos en los 8 años posteriores a la ley de AA y el periodo comparable de los 8 años previos (Gráfico 3).

**Gráfico 3.** Tipo de acceso de los artículos encontrados para investigadores del CONICET en los ocho años anteriores y posteriores a la ley de acceso abierto

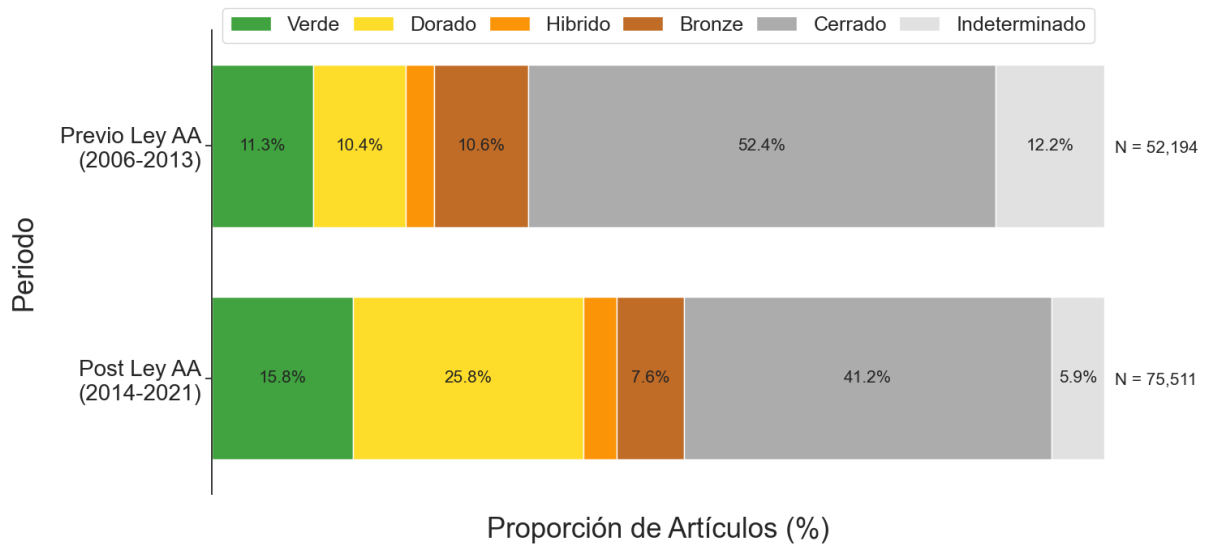

Fuente: Base de datos propia basada en datos extraídos de Open Alex en agosto de 2022 (N = 127,705)

Como se mencionó, aunque la proporción de artículos en acceso abierto, tanto por la vía verde, la dorada, híbrida y bronce, son superiores en el periodo posterior a la ley -se identificaron 75.511 para ese período y 52.194 en el periodo anterior-, se observa que esta tendencia a la apertura existía previamente. Para evaluar entonces el impacto de la ley, modelamos la proporción de artículos depositados en un repositorio argentino —que, como dijimos, es lo que se estipula por mandato— usando una regresión descrita en la ecuación (1). En ella, la variable dependiente $Y$ representa un indicador binario, codificado como 1 si el trabajo fue depositado en un repositorio argentino y 0 si no lo fue; la variable $T$ es una variable continua que indica el número de semanas transcurridos desde el comienzo del periodo previo a la ley (1 de enero, 2006); $D$ es una variable ficticia codificada como 0 para las observaciones previas a la ley y 1 para las después de la ley (1 de enero, 2014); y $P$ es una variable continua que indica el número de semanas transcurridas después de la ley. $P$ es codificada como 0 para el periodo previo a la ley. Se hizo un recorte al 31 diciembre del 2020 dado que hay una caída en el número de depósitos observados para el 2021, probablemente a causa de un retraso en depositar o en actualizar los repositorios). El resultado de está regresión muestra que la ley ha llevado a un incremento de un 5% ($p < 0.001$) en la proporción de artículos depositados semanalmente por encima de lo estimado en base a la tendencia existente, la cual mantiene la misma pendiente.

**Ecuación 1**

$$Y = \beta_0 + \beta_1 T + \beta_2 D + \beta_3 P + \delta_{md} + \varepsilon$$

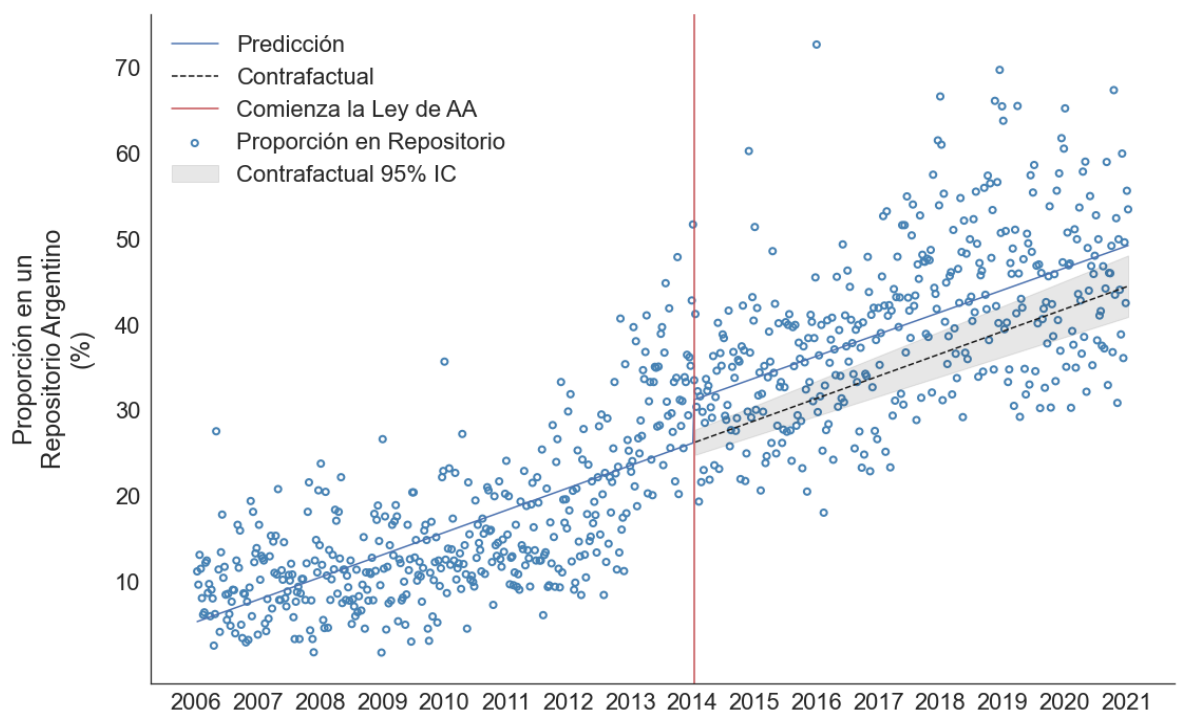

**Gráfico 4.** Proporción de artículos depositados en repositorios argentinos por semana (2006-2020)

Fuente: Base de datos propia basada en datos extraídos de Open Alex en agosto de 2022.

### *5.4. Identificación de disciplinas más abiertas (post-ley)*

En cuanto a la penetración del acceso abierto a partir de la implementación de la ley (2014-2021), en las grandes áreas temáticas definidas por el organismo puede observarse (gráfico 5), que son las Ciencias Agrarias, de Ingeniería y de Materiales (CAIM) las más propensas a publicar sus trabajos de manera cerrada (52%) mientras que las Ciencias Exactas y Naturales (CEN) y las Ciencias Biológicas y de la Salud (CBS) se mantienen cerca del promedio (40%). Solo en las Ciencias Sociales y Humanidades (CSH) el predominio es la publicación en abierto, siendo la disciplina que cuenta con mayor disponibilidad en la vía dorada (50%). Las otras ciencias le siguen muy por debajo: CBS, 28%; CAIM, 22% y las CEN, 17%. Destaca, asimismo, que la mayor disponibilidad a través de la vía verde se encuentra en las CEN con 27%, seguido por CAIM (14%), CBS (12%) y CSH (6%). La vía híbrida se mantiene siempre baja -entre 2% y 4%- en todas las disciplinas.

**Gráfico 5.** Distribución porcentual de los artículos por gran área disciplinar de CONICET según tipo de acceso: Periodo 2014-2021

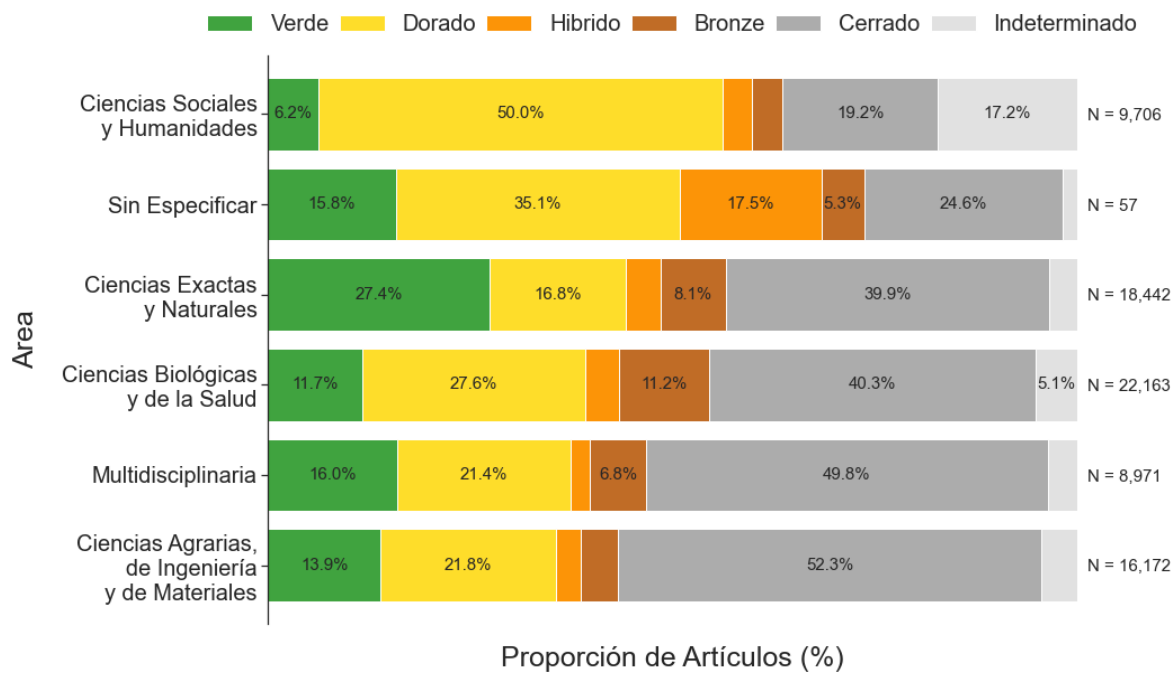

Fuente: Base de datos propia basada en datos extraídos de Open Alex en agosto de 2022 (N = 75.511)

Las grandes áreas temáticas no son homogéneas hacia su interior. En el gráfico 6 se pueden ver las variaciones de las disciplinas específicas. En el total de la serie (2014-2021) puede verse como los artículos tras las barreras de pago son la mayor parte en casi todas las disciplinas salvo por las de CSH en donde predomina en las múltiples disciplinas la vía dorada. Dos disciplinas específicas que destacan son: Hábitat y Diseño donde la vía dorada tiene mayor proporción (60%) y Astronomía y Matemáticas, disciplinas dónde la vía verde supera ampliamente a la dorada. Asimismo, puede verse que en disciplinas como la Química y la Ingeniería de Procesos el porcentaje de publicaciones cerradas es superior al 60%. Destacan por su mayor presencia de acceso híbrido Filosofía (8%), Física (7%) y Bioquímica y Biología Molecular (7%).

**Gráfico 6.** Distribución porcentual de los artículos por gran área disciplinar y disciplina específica del CONICET (2014-2021)

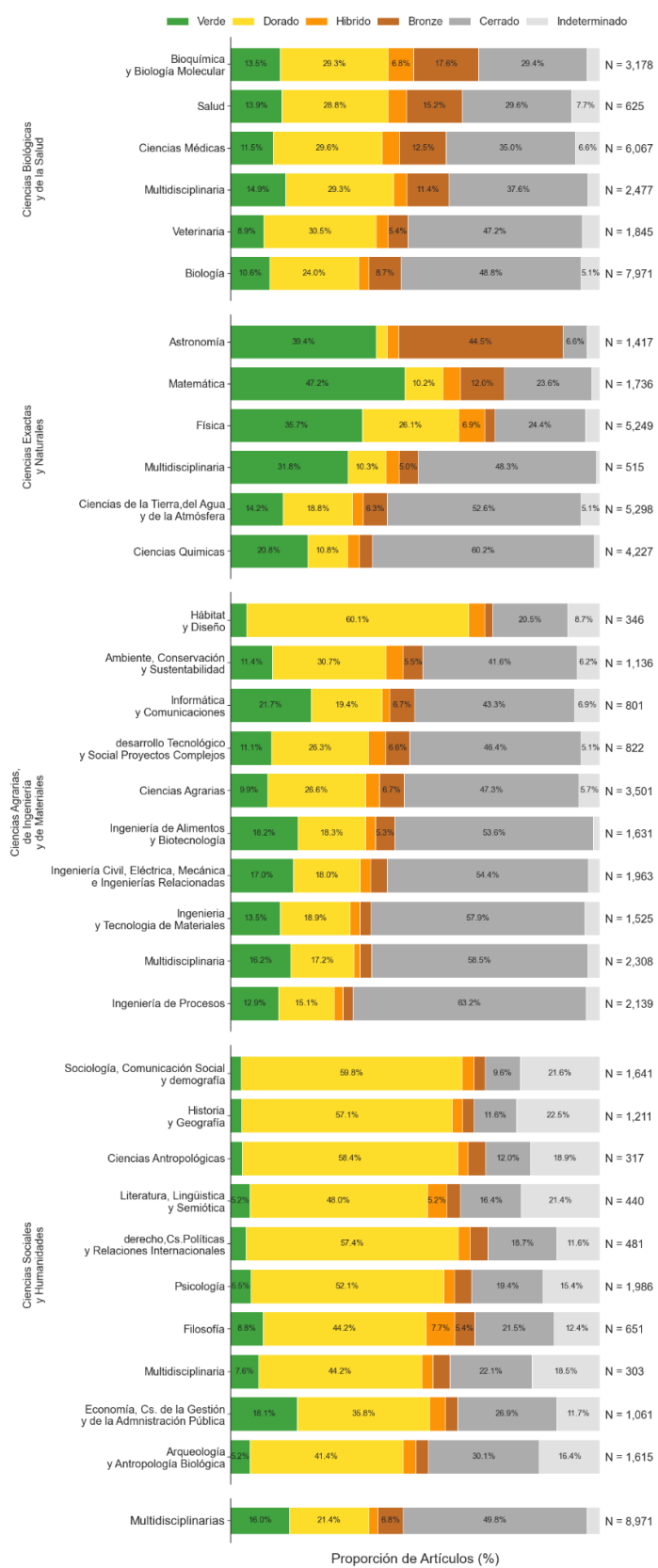

Fuente: Base de datos propia basada en datos extraídos de Open Alex en agosto de 2022 (N = 75.511)

## 6. Discusión

La aplicación de la metodología propuesta para el caso de Argentina permite detectar que la cantidad de trabajos disponibles en acceso abierto ha aumentado en los periodos estudiados, siendo la vía dorada la que presenta subas más notorias. Mientras que los repositorios—la vía verde impulsada por la legislación nacional— aportan menor proporción de trabajos solo disponibles en AA por esta vía. Los resultados obtenidos son consistentes en general con el reciente estudio de Velez Cuartas et al. (2022) en el que analizan publicaciones con al menos una afiliación argentina, aunque los porcentaje de trabajos en cada vía varían con respecto a este estudio debido a diferencias en la metodología aplicada.

Asimismo, la disponibilidad de trabajos en AA en el CONICET mantiene valores similares a los reportados en otros estudios a nivel global. Por ejemplo, un estudio que utiliza la fuente Unpaywall detectaba un 45% en acceso gratuito para el año 2015, principalmente a través de la vías dorada y bronce (Piwowar et al., 2018). Otros estudios más recientes también presentan valores similares. Uno de ellos calcula la mediana mundial de publicaciones en acceso abierto en 43% para el período 2014-2017 (Robinson-Garcia, Costa, van Leeuwen, 2020). Aunque la proporción en AA de investigadores CONICET, es mucho más baja que en las universidades de "mayor rendimiento" para las que se detecta, en el periodo 2016-2018, un 80-90% de publicaciones en abierto (Huang et al., 2020). Ambos estudios coinciden en determinar un papel fundamental a la vía verde y a las políticas nacionales que repercuten directamente en el comportamiento de las instituciones asociado con un gran aumento del nivel de acceso abierto. Incluso, Huang et al. (2020) concluyen que el crecimiento del acceso abierto en Europa y Norteamérica es impulsado por los repositorios como consecuencia de los mandatos mientras que en América Latina y África el acceso abierto crece más de la mano de la vía dorada.

Al respecto, en Argentina puede verse que mientras el acceso abierto dorado tiene un notorio aumento, la vía verde aumenta más tímidamente, aportando un valor máximo de 19% en 2018, con algunas fluctuaciones y bajas al final de la serie que, como se dijo, podrían deberse a las demoras en la carga del repositorio CONICET Digital. Sin embargo, se ve un crecimiento significativo en el número de trabajos encontrados en repositorios que ya están disponibles en revistas de AA. Aunque habría que confirmarlo para el caso de investigadores del CONICET, Hobert et al. (2021) encontró que, en Alemania entre el 2010-2017, es más probable que un artículo de una revista de AA dorado se archive en un repositorio que un artículo de otra vía de AA (híbrido, con embargo). El solapamiento de trabajos en la vía verde con la dorada y bronce muestra que los repositorios cumplen, entre otras, la función de reunir la producción nacional ya disponible en otras modalidades de acceso abierto.

Aquí cabe resaltar que nuestro análisis no nos permite atribuir está tendencia ascendente en el número de trabajos accesibles en repositorios directamente a la promulgación de la ley de AA, ya que esta tendencia era clara en los años previos. Sin embargo, a pesar de que sólo se midió un incremento de un 5% en la proporción de depósitos se pudo observar el uso de los repositorios argentinos para facilitar el creciente volumen de trabajos. La ley ha cumplido un rol importante en la creación de repositorios que se desarrollan principalmente con herramientas de fuentes abiertas (Fushimi et al., 2021). Este tipo de infraestructuras públicas y abiertas sin fines de lucro, con un enfoque de acceso abierto orientado hacia el conocimiento como bien común se alinean con los requerimientos internacionales de la ciencia abierta (Becerril-García y Gónzalez, 2021).

Este apoyo de los estados a través de diferentes organismos de ciencia y técnica nacionales para el mantenimiento de una infraestructura nacional es fundamental ante un panorama en el cual la vía dorada—la que más ha crecido— requiere cada vez de mayor financiamiento por las tasas de publicación. Aunque estos costos (APCs) aún no son comunes en las revistas latinoamericanas porque son sostenidas principalmente por fondos públicos y en base a la creación de infraestructuras con recursos compartidos (Córdoba González, 2021), sí es cada vez mayor en las editoriales comerciales internacionales (Butler et al., 2022). En este sentido Vélez Cuartas et al. (2022), proyectaron para los investigadores del CONICET un pago estimado en USD $3.602.627 en costos de APCs para el periodo 2013-2020. Tal como advierte Alperin (2022), el creciente uso de los APCs es alarmante para las revistas de latinoamericanas y requiere una respuesta de política pública, para la cual los repositorios nacionales podrían ser un componente importante.

Por supuesto, cualquier respuesta debe tomar en cuenta las diferencias disciplinares que quedan resaltadas en el presente trabajo. Al igual que estudios realizados a nivel global, se coincide en encontrar la mayor proporción de AA para las Ciencias Biomédicas y Matemáticas y las proporciones más bajas en Ingeniería y Química (Archambault et al., 2014; Piwowar et al., 2018). Sin embargo, nuestro estudio difiere en lo que hace a Humanidades para la cual en el caso de estudio se encontró un alto porcentaje de documentos disponibles en la vía dorada. Está discrepancia podría deberse a que la fuente utilizada para este estudio tiene como antecesor a la base del MAG, recurso destacado por tener una mayor cobertura en estas disciplinas que las fuentes de los otros estudios (Huang et al., 2020), pero también se podría atribuir a diferencias en el comportamiento de los investigadores de la región. Por ejemplo, para las CSH, en donde pudo verse que las publicaciones cerradas representan el 20%, los valores de las vía dorada son similares a los detectados para 2017 en Scopus (Bongiovani y Miguel, 2019). Referente al predominio de la vía dorada frente a otras áreas, deben ponerse en relación dos cuestiones entrelazadas: La primera, que el promedio de publicación de los investigadores del CONICET de estas disciplinas es realizada mayormente en publicaciones nacionales y en segundo término en revistas latinoamericanas (Beigel y Gallardo, 2021). La segunda, que las revistas de CSH editadas en el país son más numerosas que en otras disciplinas, incluso son publicadas en su mayoría por universidades nacionales (Salatino, 2019).

Las diferencias regionales y en las fuentes lleva a la necesidad de explorar, tal como lo hemos hecho en este estudio, el uso de bases de datos que capturen a las revistas y repositorios en donde se encuentra la producción Argentina. Dada la publicación en revistas nacionales, OA es una fuente prometedora (Khanna et al., 2022). Nuestro caso muestra que el uso de OA y el *script* desarrollado permitirían monitorear el avance del AA teniendo un panorama a nivel nacional sin la necesidad de pagar por bases de datos comerciales. Además, a diferencia de otros estudios realizados, este trabajo configuró el corpus de trabajos analizados utilizando como fuente a las personas, lo que podría aportar una manera de obtener información más actual y precisa, aunque la rigurosidad de OA aún debe ser confirmada de manera empírica.

### 6.1. Limitaciones del estudio

La falta de conocimiento sobre la calidad de los datos de OA es una limitación importante de este estudio. Nuestra intención fue demostrar el potencial de esta fuente y desarrollar unos *scripts* y metodología para monitorear el AA de un conjunto de investigadores que pueda ser aplicada de manera fácil a otras comunidades. El presente trabajo ha demostrado este potencial, pero se requiere cautela al interpretar los datos dada la novedad de la fuente.

Asimismo, será necesario la realización de otros estudios para validar la metodología de búsqueda por nombre y los filtros utilizados a fin de maximizar la cantidad de trabajos recuperados y minimizar los trabajos erróneos obtenidos.

Dentro de los datos que se lograron obtener, otra limitación del estudio y de la fuente está dada por los datos de la disponibilidad en acceso abierto de los artículos que parece depender en gran medida a que tengan un DOI asignado. Otra limitación está relacionada al bajo número de artículos pertenecientes a las CSH. Aunque la cantidad parece ser mayor al de otras fuentes, sigue siendo baja en comparación con la de otras grandes áreas de conocimiento. Aún así, se considera que los resultados son aceptables. Las limitaciones de este estudio son, a nuestra opinión, comparables con las de otros estudios que dependen de fuentes más acotadas que OA.

## 7. Conclusiones

El estudio realizado permite decir que actualmente al menos un 41% de los artículos identificados para investigadores del CONICET están disponibles en alguna modalidad de acceso abierto. Asimismo, se evidenció que tanto la vía verde como la dorada han tenido una evolución positiva en los últimos 20 años tal cómo lo muestra la tendencia internacional. El recorte realizado considerando la implementación de la legislación nacional (2014-2021) mejora la proporción de trabajos disponibles en acceso abierto llegando a alcanzar el 46% de los artículos. De acuerdo con los datos relevados para este estudio se puede decir que el avance del acceso abierto en Argentina está impulsado más rápidamente por las revistas doradas que por la inclusión de trabajos publicados en títulos cerrados en repositorios abiertos institucionales, a pesar de que la legislación requiere esta modalidad. Se pudo determinar al analizar la proporción de depósitos semanales en un repositorio, que en el caso de la producción relacionados con los investigadores de CONICET la ley ha llevado a un incremento de un 5%.

Es preciso destacar la importancia de contar con fuentes completamente abiertas y gratuitas que agreguen información científica de todas las latitudes y ofrezcan servicios para acceder a sus datos sin mayores complicaciones. Este estudio ha mostrado que es posible dar seguimiento a la producción científica de un sistema nacional con una fuente abierta y gratuita, como es OA. Esta posibilidad se podría expandir si fuese posible también contar con los datos de ciencia y técnica oficiales (personal, recursos, publicaciones, etc.) de manera abierta y actualizada. Esto, junto con el control de autoridades de los nombres de los investigadores, tarea a realizarse por las bibliotecas nacionales de los países, simplificaría la metodología propuesta resolviendo inconvenientes con los nombres y permitiendo la validación de la información de manera más precisa.

La importancia de la ley de AA de Argentina quizás se pueda entender mejor al verla dentro del contexto internacional, en el cual el AA con pago de APCs es cada vez más común. La ley actual ha creado apoyos para la vía verde con la creación de repositorios, pero se puede observar que los investigadores de CONICET han adoptado la vía dorada. Al respecto sería adecuado acompañar el diseño y la implementación de políticas nacionales, regionales e internacionales que fomenten la edición de revistas bajo la llamada modalidad diamante (ni autor, ni lector pagan) a fin de favorecer prácticas en abierto en todas las disciplinas. Esto puede hacerse a la par de políticas, como la ley de AA de Argentina, que incentiven el autoarchivo aunque deberán también acompañarse por cambios en la evaluación de las trayectorias científicas.

# Bibliografía